\documentclass[]{emulateapj}
\usepackage{graphicx}
\usepackage[suffix=]{epstopdf}
\usepackage{natbib}
\usepackage{amsmath}
\usepackage{hyperref}

\shortauthors{Davenport et al.}
\shorttitle{Near IR Variability of AGN}

\begin{document}

\title{SDSSJ14584479+3720215: A Benchmark $JHK_s$ Blazar Light Curve \\from the 2MASS Calibration Scans}

\author{James R. A. Davenport\altaffilmark{1,2},
John J. Ruan\altaffilmark{2},
Andrew C. Becker\altaffilmark{2},
Chelsea L. Macleod\altaffilmark{3},
Roc M. Cutri\altaffilmark{4}}

\altaffiltext{1}{Corresponding author: jrad@astro.washington.edu}
\altaffiltext{2}{Department of Astronomy, University of Washington, Box 351580, Seattle, WA 98195, USA}
\altaffiltext{3}{Institute for Astronomy, University of Edinburgh, Edinburgh EH9 3HJ, U.K.}
\altaffiltext{4}{Infrared Processing and Analysis Center, California Institute of Technology, Pasadena, CA 91125, USA}

\begin{abstract}
Active galactic nuclei (AGNs) are well-known to exhibit flux variability across a wide range of wavelength regimes, but the precise origin of the variability at different wavelengths remains unclear. To investigate the relatively unexplored near-IR variability of the most luminous AGNs, we conduct a search for variability using well sampled $JHK_s$-band light curves from the 2MASS survey calibration fields. Our sample includes 27 known quasars with an average of 924 epochs of observation over three years, as well as one spectroscopically confirmed blazar (SDSSJ14584479+3720215) with 1972 epochs of data. This is the best-sampled NIR photometric blazar light curve to date, and it exhibits correlated, stochastic variability that we characterize with continuous auto-regressive moving average (CARMA) models.  None of the other 26 known quasars had detectable variability in the 2MASS bands above the photometric uncertainty. A blind search of the 2MASS calibration field light curves for AGN candidates based on fitting CARMA(1,0) models (damped-random walk) uncovered only 7 candidates. All 7 were young stellar objects within the $\rho$ Ophiuchus star forming region, five with previous X-ray detections. A significant $\gamma$-ray detection (5$\sigma$) for the known blazar using 4.5 years of Fermi photon data is also found. We suggest that strong NIR variability of blazars, such as seen for SDSSJ14584479+3720215, can be used as an efficient method of identifying previously-unidentified $\gamma$-ray blazars, with low contamination from other AGN.
\end{abstract}


%%%%%%%%%%%%%%%%%%%%%%%%%%%%%%%%%%%%%%%
\section{Introduction}

The temporal flux variability from active galactic nuclei (AGN), detectable in nearly all wavelength regimes, contains information on the underlying emission processes and source geometry that is otherwise difficult to probe \citep{ulirch1997}. However, precise details of the physical mechanism generating the observed nuclear variability in AGN remain unclear \citep{antonucci2013}. Current and future large-scale photometric time-domain surveys have motivated many recent studies of the optical broadband variability properties of various AGN subclasses using large numbers of well-sampled light curves. This has been especially useful for AGN identification and selection \citep{kelly2009,kozlowski2010a,zu2013}. 

Beyond the optical, large-scale surveys of AGN variability have been pursued at many other wavelengths, including the radio \citep[][]{thyagarajan2011}, ultraviolet \citep[][]{gezari2013}, and $\gamma$-ray regimes \citep[][]{ackermann2012}. Fewer studies of AGN variability have focused on the infrared (IR), due in part to the expectation that non-variable dust and the host galaxy dominate the emission at these wavelengths \citep[e.g. see][]{kishimoto2008}. Explanations of AGN variability in the optical as due to localized temperature fluctuations in inhomogeneous accretion disks \citep{ruan2014} or global accretion rate changes \citep{pereyra2006} predict little flux variability from disk emission in the IR. Reprocessing of variable optical emission by a dusty torus has been suggested by \citet{suganuma2006} as a possible cause for the NIR variability, although the time-lags between the optical and NIR are difficult to detect. For example \citet{mchardy2007} have shown small time lags between NIR and X-ray data for the AGN 3C273, with the NIR variability leading the X-ray by $\sim$1.5 days, implying this NIR variability originates near the nucleus.

A study of NIR variability in the Spitzer Deep Wide-field Survey Bo\"{o}tes field by \citet{kozlowski2010b} showed that although only 1.1\% of objects appear to be variable in the near-IR (NIR), the vast majority of the variable objects were AGN.  Small samples of AGN monitored in both the optical and NIR have shown that while nuclear variability is prevalent, the amplitude of variations  decreases towards longer wavelengths. For example, \citet{honig2011} have produced realistic models of the optical to NIR variability for NGC 4151, which show a small amplitude time lag and decreasing amplitude variability as a function of wavelength. However, this is not true for blazars, whose NIR flux is dominated by non-thermal emission from a relativistic jet, rather than circum-nuclear  dust which is not expected to be non-variable \citep{cutri1985}. Recent results from optical/NIR monitoring of $Fermi$ $\gamma$-ray blazars have shown that flat-spectrum radio quasars are more variable in the NIR than in the optical \citep[][]{bonning2012,sandrinelli2013}. \citet{sandrinelli2014} have used 7 year photometric light curves of 7 blazars, spanning the optical to NIR bands ($VRIJHK$), to find that variability amplitudes increase with wavelength for these objects.

In this study we search for NIR variability of luminous AGN using well sampled light curves from the Two Micron All Sky Survey (2MASS) calibration data, finding one highly variable blazar (SDSSJ14584479+3720215). We explore the utility of NIR variability-based selection of AGN, including its use in identifying counterparts to $Fermi$ $\gamma$-ray sources. We also study the nature of non-AGN contaminants with similar NIR colors and variability properties, laying groundwork for the AGN variability science possible with current and future multi-epoch IR surveys. 

%%removed per referee's comment 
%The structure of this letter is as follows: in \S2 we discuss the selection of our AGN sample in the 2MASS calibration fields. In \S3 we investigate the NIR light curves of our AGN sample, and model those with NIR variability. In \S4 we discuss contamination by other variable objects in our sample and their implications on AGN selection using NIR variability. In \S5 we discuss how NIR variability properties can inform and drive AGN selection in other wavelengths by discovering $\gamma$-ray emission from an AGN in our sample using its NIR variability properties. Finally in \S6 we place this work in the context of future IR time-domain surveys.

%% NOTE: wise_ccd.eps newest version being made on Linux machine, not laptop!!
%%             run in : /astro/users/jrad/research/2mass_cal/drw/
%%             code in: ~/Dropbox/jrad_idl/2massscripts/agn/agnplots.pro
\begin{figure}[]
\centering
\includegraphics[width=3.5in]{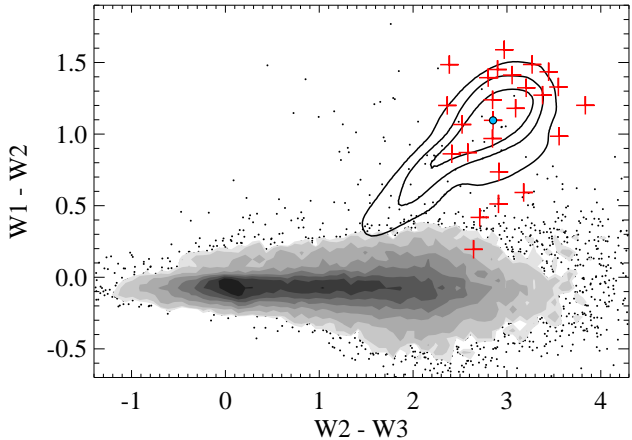}
\caption{WISE color-color space for the Cal-PSWDB sources with WISE detections (black points and grey filled contours), known AGN (red crosses), and the WISE blazar strip (black open contours) from \citet{choi2014}. The known blazar SDSSJ14584479+3720215 is highlighted (blue circle).}
\label{wcolor}
\end{figure}

%%%%%%%%%%%%%%%%%%%%%%%%%
\section{Data and AGN Selection}
The 2MASS survey observed the full sky in the near-IR over the timespan of June 1997 to February 2001 using the $J$, $H$, and $K_s$ bands \citep{2mass}. Photometric zero-points for calibrating the survey were based on hourly observations of a set of 40 standard fields, each $8\farcm5\times1^\circ$ in size, and spaced evenly throughout the sky. These fields were visited repeatedly over the course of the survey, resulting in 562 to 3692 epochs of observation per field. This produced $JHK_s$ light curves for 113,030 individual objects, known as the 2MASS Calibration Point Source Working Database \citep[hereafter Cal-PSWDB,][]{cutri2006,plavchan2008a}. These light curves are the most precise NIR time domain survey to date and provide a novel dataset in which to study the infrared variability properties for a wealth of astrophysical phenomena. This dataset has produced the best sampled NIR light curve for a RR Lyr star \citep{szabo2014}, a hunt for stellar flares at long wavelengths \citep{davenport2012}, 
a large sample of young stellar objects \citep[YSO's,][]{plavchan2008b,parks2014}, and many well characterized binary star systems \citep[][Davenport 2014 in prep]{becker2008,quillen2014}.
Here we make use of these data to search for NIR variability of AGN with unprecedented precision.

The Wide-field Infrared Survey Explorer \citep[WISE,][]{wise} has created the deepest mid-IR survey of the entire sky in four photometric bands, from 3.4 $\mu$m to 22 $\mu$m. Several recent studies have demonstrated the utility of WISE colors for effectively separating AGN from stars and unresolved galaxies. We spatially matched the Cal-PSWDB objects to the WISE all sky data release using a match radius of 1\farcs5. This resulted in 52,148 objects with $JHK_s$ light curves and a detection in at least one WISE filter. The distribution of these point sources in WISE color--color space is shown in Figure \ref{wcolor}. For comparison, we show the WISE ``blazar strip'', defined in \citet{choi2014} as the Gaussian Kernel Density Estimate of the WISE colors of blazars from \citet{massaro2009}.

To find previously known AGN in our data, we spatially matched these 52K objects to the ``Milliquas Catalog,'' version 3.3\footnote{\href{http://quasars.org/milliquas.htm}{http://quasars.org/milliquas.htm}}. We recovered 27 known AGN with light curves in Cal-PSWDB, 25 of which had WISE colors shown in Figure \ref{wcolor}. These AGN had between 26 and 3482 epochs of Cal-PSWDB photometry, with an average of 924 epochs. Of these, 19 have been spectroscopically confirmed by the Sloan Digital Sky Survey \citep[SDSS;][]{york2000,bovy2011,paris2012}, one of which has been identified as a blazar (SDSSJ14584479+3720215).

%%%%%%%%%%%%%%%%%%%%%%%%%
\section{AGN Variability in the NIR}

%% NOTE: blazar_lc.eps newest version being made on Linux machine, not laptop!!
%%             run in : /astro/users/jrad/research/2mass_cal/drw/
%%             code in: ~/Dropbox/jrad_idl/2massscripts/agn/agnplots.pro
\begin{figure}[!t]
\centering
\includegraphics[width=3.5in]{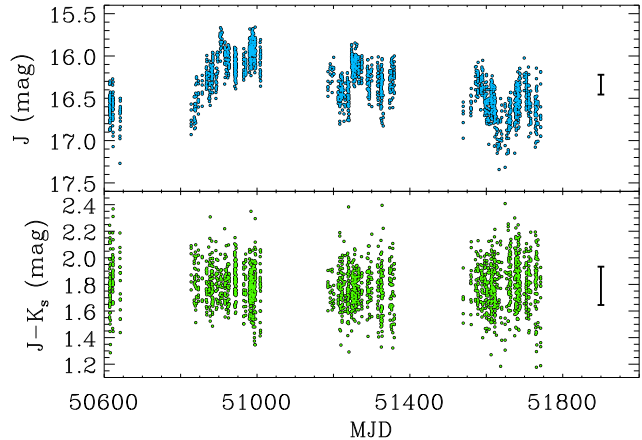}
\caption{$J$-band (top) and $J-K_s$ color (bottom) light curves for the known blazar, SDSSJ14584479+3720215. The median photometric error is shown for comparison (vertical bar).}
\label{lc}
\end{figure}

\vspace{.05in} %% keep caption text from running in to body

This set of known AGN with well sampled Cal-PSWDB light curves creates a unique dataset in which to search for AGN variability in the NIR. To determine if these objects were variable in the 2MASS bands, we computed the error-weighted root mean square variability for each light curve, $\Sigma$, following the prescription in \citet{sesar2007}. Objects were classified as variable if they had $\Sigma_J>10^{-2}$ mag. Of the ``Milliquas'' AGN in our dataset, only the known blazar, SDSSJ14584479+3720215, exhibited variability above this threshold. The $J$-band and $J-K_s$ color light curves for this blazar, with 1972 epochs of Cal-PSWDB photometry spanning over three years, are presented in Figure \ref{lc}. The remaining 26 known AGN in our sample displayed no significant NIR variability using this metric.

We found strong variability in SDSSJ14584479+3720215 in all three 2MASS bands with our densely sampled light curves. However, as seen in Figure \ref{lc} no significant variation in $J-K_s$ color was found for this object, indicating the underlying variability was ``grey'' over this wavelength regime. A larger sample of such well studied variable objects is needed to test if this is truly a generic property of blazars in the NIR.

Characterizing photometric variability timescales has proven to be a very efficient means of selecting a clean sample of AGN from time domain surveys \citep[e.g.][]{kozlowski2010a, macleod2011, butler2011}. The so-called ``damped random walk'' (DRW) model provides a robust fit to AGN light curves as a function of two observable free parameters: the characteristic timescale $\tau$, and the variability amplitude SF$_\infty$. Stars are easily removed when fit with this model, as their primary sources of variability are typically stochastic (e.g. flares or flickering) or strongly periodic (such as pulsations or rotation) in nature. Combining DRW variability models with multi-wavelength photometry provides even greater power in separating blazars from ``normal'' AGN \citep{ruan2012}.

From our Cal-PSWDB--WISE matched sample we selected all light curves having at least 50 good $J$-band epochs (PH\_QUAL=A,B or C), median $J$-band magnitudes brighter than 16.5 mag, and WISE colors of $W1-W2 > 0.3$ to remove main sequence stars \citep{davenport2014}. We also required sources to be detected as variable using the $\Sigma_J>0.01$ cut described above, and removed objects known to be periodic (binaries and pulsating variables) or long period variable stars \citep[][Davenport 2014 in prep]{plavchan2008b}. This yielded 22 variable Cal-PSWDB targets to search for DRW timescales, which included our known blazar object, but did not select any other previously identified AGN. For completeness we also analyzed the other 26 known ``Milliquas'' sources for DRW variability.

For each object we down-sampled the light curves to a single data point per night, using the median of the $J$-band photometry, which reduced the light curves to an average of 144 epochs. We then ran the DRW fitting code from \citet{kozlowski2010a} and \citet{macleod2010} on the reduced $J$-band light curves for all 47 objects. We used the probability thresholds outlined in \citet{macleod2010} to determine which light curves had significant DRW variability. The known blazar had a weakly constrained DRW timescale of $\log \tau = 2.0^{+4.0}_{-1.9}$ days, and a significant driving amplitude of $\log \sigma = -0.11^{+0.09}_{-0.2}$ mag year$^{-1/2}$. The relative likelihood for the DRW versus an infinite or unconstrained timescale was 
log(L$_{like}$/L$_{inf}$) =1.4, where values greater than ~0.05 are considered significant. Similarly, the relative likelihood for the DRW model versus pure noise was 
log(L$_{like}$/L$_{noise}$) = 153, which was much greater than the threshold for significance of 0. None of the other 26 previously identified AGN showed any signs of significant DRW variability. However, 7 of the variability-selected objects did show significant DRW variability. We discuss the origin of these objects in the following section.

We additionally fit these light curves using the Continuous Auto-Regressive Moving Average, or CARMA(p,q), models of \citet{kelly2014}, where $p$ is the auto-regressive order and $q$ the moving average order. The DRW is a special case of the CARMA models, equivalent to a CARMA(1,0), or a first-order continuous auto-regressive process with no moving average. In the CARMA formalism, a DRW model's power spectral density is described as a single Lorentzian function with a characteristic break frequency. As mentioned above, the DRW has been shown to empirically model the behavior of QSO light curves well, but is otherwise inflexible. By looking at higher order CARMA models we search for more complex behavior in the light curve power spectral density, such as stochastic variability on multiple timescales or quasi--periodic oscillations. For SDSSJ14584479+3720215 we found a moderate preference for the CARMA(2,0) model compared to a DRW when using the deviance information criterion for model selection. The DRW model provides a sufficient fit to the data, in that the sequence of residuals is independently and normally distributed. A Markov Chain Monte Carlo analysis of the CARMA(1,0) model provided 1-$\sigma$ confidence levels on the timescale of $31 < \tau < 115$ days, and on the standard deviation of the driving noise of between 0.14 and 0.20 magnitudes.

%%%%%%%%%%%%%%%%%%%%%%%%%%%%%
\section{YSO Contamination} 
The recovered DRW timescales for the seven variability-selected objects we found in the previous section had an average timescale of $\log \tau = 1.05$ days, significantly shorter than the typical timescale seen in the optical for AGN \citep{macleod2010}, as well as for our blazar target. This was reproduced in the CARMA(1,0) models as well. These seven objects all reside in field \# 90009, which was centered on the star forming region $\rho$ Ophiuchus, making these likely YSOs. Of these YSOs, five had X-ray detections in the literature, one coming from a targeted X-ray monitoring campaign of the field \citep{gagne2004}, and four from the Chandra X-ray source catalog \citep{evans2010}. The other two YSOs in our sample did not have a published X-ray counterpart within 0.1$^\circ$. NIR light curve properties for YSOs within this field in the Cal-PSWDB have been studied in great detail by \citet{plavchan2008b} and \citet{parks2014}.

These seven YSOs had mid-IR colors that placed them within the blazar strip, NIR variability amplitudes comparable to our known blazar, and light curves best parameterized by a CARMA(1,0) model. We thus highlight YSOs as a potentially important source of contamination for classifying AGN from future multi-wavelength time domain searches. While using longer wavelength filters can improve the efficiency of selecting AGN from normal stars \citep[e.g.][]{massaro2012}, YSOs remain a potential contaminant for object classification \cite[see also color spaces in Fig 7 of][]{koenig2012}. However, these objects may be distinguished from bona fide AGN based on their $\sim$10X shorter characteristic timescales under a CARMA(1,0) model. YSOs are also mostly confined to the Galactic disk and dense star forming regions, and as such can be avoided by surveys.

%%%%%%%%%%%%%%%%%%%%%%%%%%%%%
\section{A $\gamma$-ray detection of SDSSJ14584479+3720215} 

\begin{figure*}[!t]\centering
\includegraphics[width=2.25in]{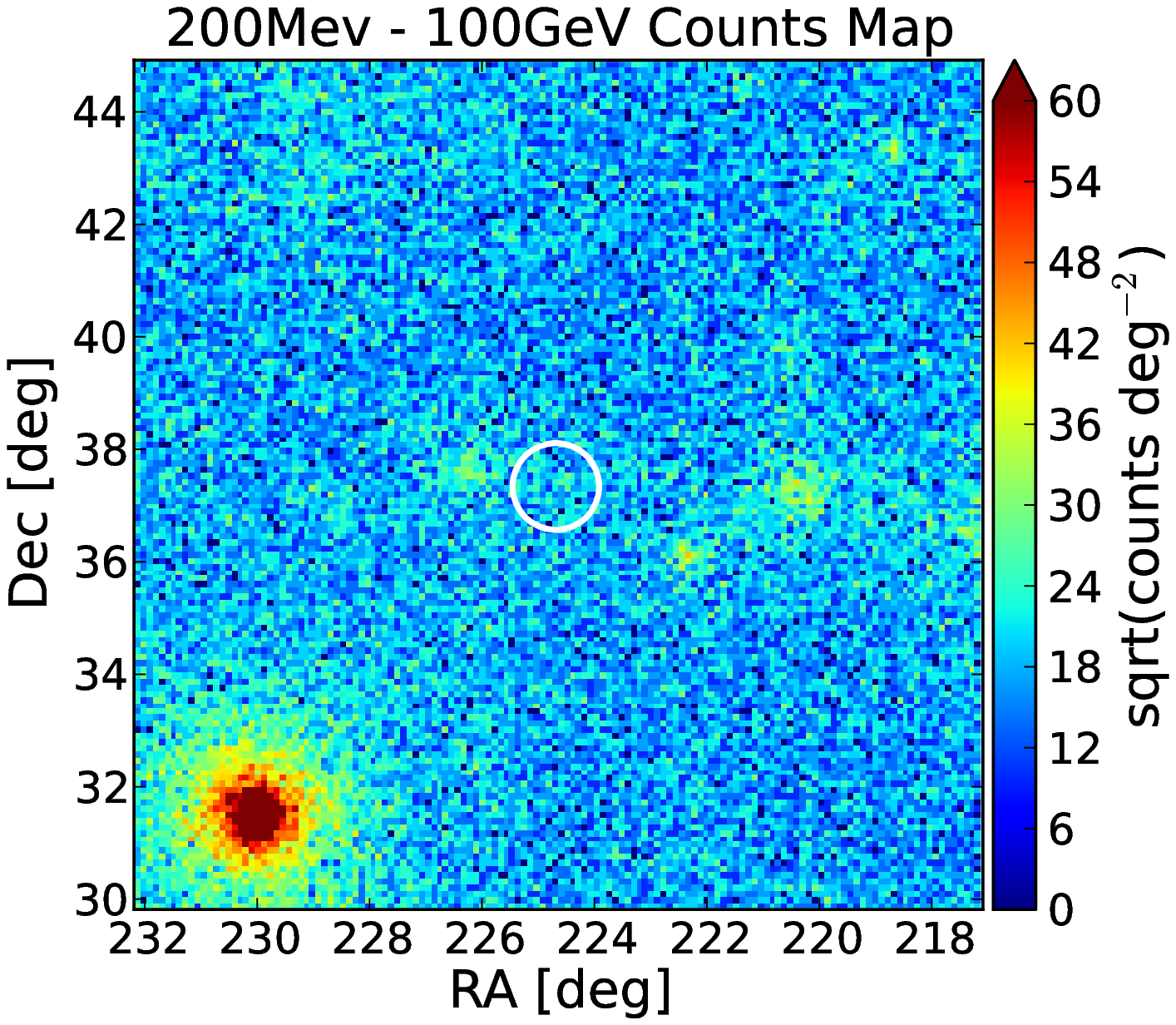}
\includegraphics[width=2.25in]{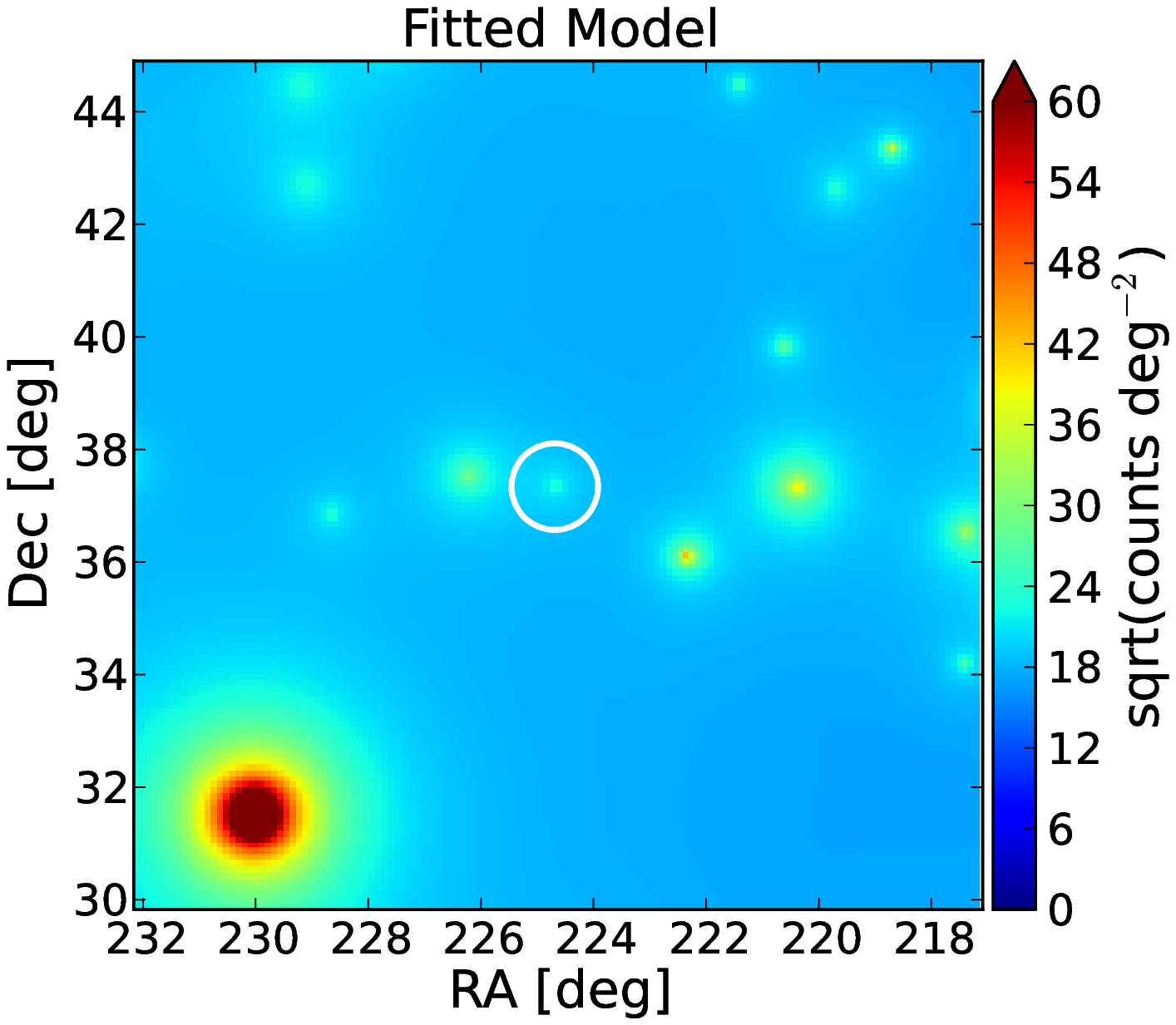}
\includegraphics[width=2.25in]{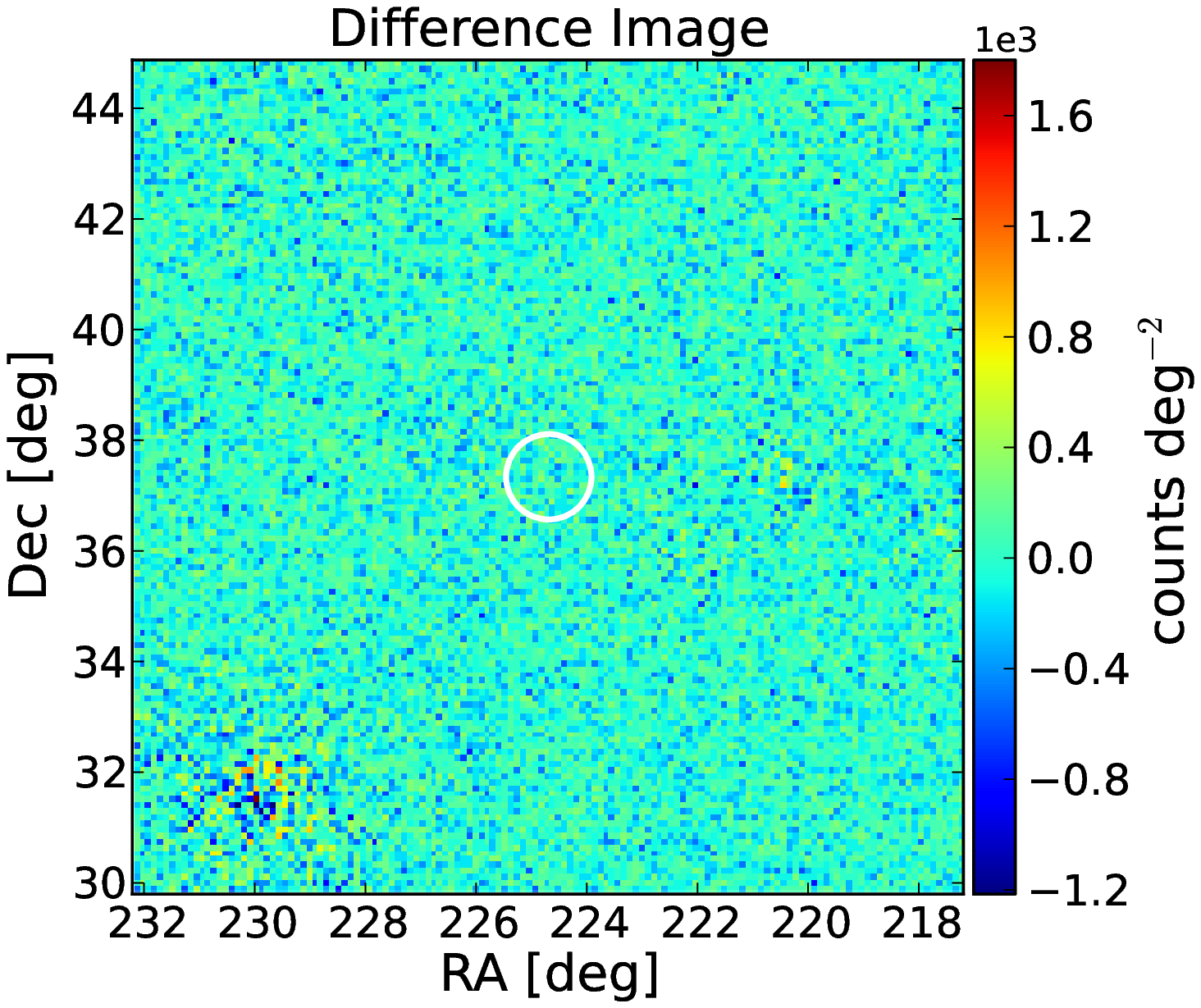}
\caption{Left: 200 MeV - 100 Gev counts maps of $\sim$4.5 years of {\it Fermi} data, centered on the known blazar (circled). Middle: Simulated counts map for the best fit model of all sources in the field, including the known blazar, detected with $TS = 30$. Right: Difference image between the observed and model counts maps.}
\label{fig:fermi}
\end{figure*}

The {\it Fermi Gamma-ray Space Telescope} provides the deepest survey to date in the 100 MeV to 100 GeV regime, and has discovered 886 $\gamma$-ray AGN in the 2nd {\it Fermi} AGN Catalog \citep{ackermann2011}. The vast majority of these AGN are blazars. However, $\sim$30\% of {\it Fermi} sources in the {\it Fermi} 2nd Point Source Catalog \citep[2FGL;][]{nolan2012} are unidentified, largely due to the poor angular resolution of {\it Fermi}'s pair-conversion Large Area Telescope \citep[LAT;][]{atwood2009}. All reliably-associated AGN in the 2nd {\it Fermi} AGN Catalog are confidently detected by {\it Fermi} with Test Statistic $TS > 25$, corresponding to approximately $>5\sigma$ detections. For a {\it Fermi} counts map of a Region of Interest (ROI) centered on a new possible source, the TS detection significance is the defined to be the ratio of likelihoods TS = 2(log$L$(source) - log$L$(nosource)), where $L$(source) is the likelihood of a model of all known sources in the ROI and a new source at the central location, while $L$(nosource) is the likelihood for the same model but no new source at the central location \citep{nolan2012}. The detection significance is approximately $\sqrt{TS}$. Thousands of potential faint $\gamma$-ray blazars lie below the $TS > 25$ ($>5\sigma$) detection threshold, but are difficult to localize due to extremely low photon counts. These faint $\gamma$-ray blazars have large positional uncertainties, with error ellipses with radius $\gtrsim$20\arcmin~for 3$\sigma$ Fermi detections using 4 years of photon data \citep{ballet2013,thompson2014}.

The strong $JHK_s$ variability exhibited by the blazar SDSSJ14584479+3720215 relative to the quasars in our sample suggests that this NIR variability is likely to be indicative of strongly beamed jets, which are characteristic of blazars, and are known to produce strong $\gamma$-ray emission. This allows for NIR-variability based identification of previously unidentified or undetected $\gamma$-ray blazars. SDSSJ14584479+3720215 was not included in the {\it Fermi} 2FGL catalog, and thus had $TS < 25$ is 2 years of photon data. However, motivated by the strong NIR variability we observe, we investigated whether this blazar can be detected at high significance using more photon data. We downloaded approximately 4.5 years of available Pass 7 {\it Fermi} photon data of the Source event class, spanning Mission Elapsed Time 239557417 to 383570220 seconds. We selected photons in the 200 MeV to 100 GeV energy range, within a 10$^\circ$ Region of Interest (ROI) centered on the blazar.

Using the standard binned likelihood analysis scripts provided in the Fermi Science Tools package, we performed photon event selection with a ROI-based zenith angle cut of $<$100$^\circ$ and a rock angle cut of $<$52$^\circ$. We produced a spatial model of the likely $\gamma$-ray sources within 15$^\circ$ of the blazar based on the 2FGL catalog, using the 2FGL best fit values for the template spectral model of these sources. We included the blazar SDSSJ14584479+3720215 as an additional point source, modeled as a simple power-law spectrum. The spectral normalization for all sources within 15$^\circ$ was allowed to vary, and all spectral parameters were set as free parameters within the ROI. We included the appropriate 2-year Pass 7 Galactic diffuse emission and extragalactic isotropic diffuse emission models in our source model fit to the observed counts map. 
The blazar SDSSJ14584479+3720215 is detected with $TS = 30$, approximately a 5$\sigma$ detection, with a $\gamma$-ray power-law spectral slope of $2.40 \pm 0.18$, typical of {\it Fermi} blazars \citep{ackermann2011}. This detection significance is consistent with the $TS \simeq 30$ found in a preliminary version of the {\it Fermi} 4-year point source catalog by \citet{ballet2013}, produced using 4 years of photon data. We note minor discrepancies in our modeling of the brightest object in the field, noticeable in the lower-left corner of the difference image. These are well-known issues in {\it Fermi} source analysis caused by the finite resolution of the counts maps, and do not affect our detection.

%%%%%%%%%%%%%%%%%%%%%%%%%%%%%
\section{Discussion}
We have conducted a search for NIR variability from AGN using the 2MASS C al-PSWDB light curves. No significant variability was found for 26 of the 27 known AGN in our sample. One previously known AGN, showed strong variations in the 2MASS filters, and we have presented the 1972 epoch NIR photometric light curve for this blazar (SDSSJ14584479+3720215). This previously known blazar is a benchmark object, displaying almost 1 magnitude of variability in the $JHK_s$ bands, but with no significant variation in $J-K_s$ color. This is the best sampled NIR light curve of a blazar ever measured. We note a handful of other AGN candidates from the 2MASS Cal-PSWDB were recently identified by \citet{quillen2014}, including the blazar SDSSJ14584479+3720215.

Our detection of SDSSJ14584479+3720215 as a $\gamma$-ray source, selected by infrared variability, has implications on the association and classification of multi-wavelength counterparts of {\it Fermi} sources. Current selection methods such as mid-IR color \citep{massaro2012} and optical variability \citep{ruan2012} suffer from low efficiency. This is often due to ``normal'' Type 1 quasars, which have similar colors in the mid-IR and are also variable in the optical. As this pilot study demonstrates, these issues may be surmounted by the inclusion of IR variability information, since normal AGN are not significantly variable at this wavelength regime.

A search for other such light curves in the 2MASS Cal-PSWDB reveals a large number of contaminating YSOs. This study establishes a baseline for the AGN variability science possible with current and future multi-epoch IR missions such as the WISE, the Wide Field Infrared Survey Telescope \citep[WFIRST;][]{green2012}, and Vista Variables in the V\'ia L\'actea \citep[VVV;][]{minniti2010} surveys.

%%%%%%%%%%%%%%%%%%%%%%%%%
\acknowledgements
We thank P. Plavchan and J. Parks for helpful discussions of YSO variability and sharing an early version of their manuscript. The authors acknowledge support from NASA ADP grant NNX09AC77G and NASA NNX14AK26G. JJR is supported by NASA Fermi grant  NNX14AQ23G.

This publication makes use of data products from the Two Micron All Sky Survey, which is a joint project of the University of Massachusetts and the Infrared Processing and Analysis Center/California Institute of Technology, funded by the National Aeronautics and Space Administration and the National Science Foundation.

This research has made use of the NASA/IPAC Extragalactic Database (NED) which 
is operated by the Jet Propulsion Laboratory,California Institute of Technology, 
under contract with the National Aeronautics and Space Administration.

This research has made use of data obtained from the Chandra Source Catalog, 
provided by the Chandra X-ray Center (CXC) as part of the Chandra Data Archive.

This publication makes use of data products from the Wide-field Infrared 
Survey Explorer, which is a joint project of the University of California, Los
Angeles, and the Jet Propulsion Laboratory/California Institute
of Technology, funded by the National Aeronautics and Space
Administration.

\begin{deluxetable*}{lcccccc}
\setlength{\tabcolsep}{0.02in} 
\tabletypesize{\tiny}
\tablecolumns{7}
\tablecaption{NIR Properties of the 27 ``Milliquas'' AGN found in the Cal-PSWDB. Our benchmark blazar is indicated ($\dagger$).}
\tablehead{
	\colhead{2MASS ObjectID}&
	\colhead{N$_{\rm epoch}(J)$} &
	\colhead{$\langle J\rangle$} &
	\colhead{$\langle J-K_s\rangle$} &
	\colhead{$W1$} &
	\colhead{$W1-W2$} & 
	\colhead{$W2-W3$}
	}
\startdata
J$002410.86-015646.6$ &     1063 &    16.86 &     1.45 &    14.91 &     1.33 &     3.55\\
J$015454.88+004044.6$ &       30 &    16.96 &     1.43 &    15.80 &     1.41 &     3.05\\
J$015451.41+005933.2$ &      339 &    17.00 &     1.51 &    15.29 &     0.42 &     2.71\\
J$015429.74+002711.0$ &       86 &    17.03 &     1.52 &    15.61 &     1.27 &     3.38\\
J$034105.84+070917.4$ &       11 &    17.08 &     1.18 &    15.39 &     1.48 &     2.39\\
J$055708.14-002414.3$ &      254 &    16.99 &     1.48 &   \ldots &   \ldots &   \ldots\\
J$085116.85+120028.6$ &     2580 &    16.72 &     1.75 &    13.58 &     0.97 &     2.85\\
J$094235.96+591354.7$ &       68 &    17.08 &     1.49 &    15.22 &     1.49 &     3.27\\
J$094249.43+593206.6$ &      778 &    16.23 &     1.56 &    13.39 &     0.87 &     2.58\\
J$122125.95-001006.1$ &     1873 &    16.68 &     1.41 &    14.89 &     0.51 &     2.91\\
J$122131.10+000742.7$ &        9 &    17.01 &   \ldots &    15.20 &     1.32 &     3.20\\
J$122144.68-001141.5$ &       19 &    17.09 &     1.76 &    15.78 &     1.43 &     3.44\\
J$122152.13+001719.0$ &       49 &    17.02 &     1.41 &    15.77 &     0.59 &     3.18\\
J$121427.03+350907.9$ &     2737 &    16.29 &     1.10 &    13.38 &     1.24 &     2.85\\
J$121418.95+352920.6$ &       18 &    17.10 &   \ldots &    16.50 &     1.20 &     3.84\\
J$121408.54+355021.5$ &     1212 &    16.91 &     1.56 &    14.63 &     0.20 &     2.64\\
J$144057.24-000951.1$ &      916 &    16.40 &     1.13 &    13.80 &     1.59 &     2.97\\
J$150026.22-005428.2$ &     1568 &    16.79 &     1.64 &    14.16 &     1.18 &     3.09\\
J$150043.39-005820.7$ &     1500 &    16.50 &     1.48 &    13.88 &     0.74 &     2.92\\
J$145846.09+371235.9$ &       58 &    17.06 &     1.38 &    15.32 &     0.98 &     3.56\\
J$145844.84+372021.8^\dagger$ &     1927 &    16.33 &     1.77 &    14.03 &     1.10 &     2.85\\
J$163124.44+295301.8$ &     1191 &    15.51 &     1.73 &    12.46 &     1.07 &     2.52\\
J$204110.28-052626.3$ &       21 &    16.99 &     1.49 &   \ldots &   \ldots &   \ldots\\
J$220037.69+211051.9$ &      888 &    16.75 &     1.32 &    14.34 &     1.45 &     2.90\\
J$220028.41+203902.1$ &       15 &    16.88 &   \ldots &    15.22 &     1.20 &     2.36\\
J$231808.32+001152.3$ &        7 &    16.99 &     1.46 &    15.06 &     1.39 &     2.80\\
J$231821.13+002937.2$ &      358 &    16.95 &     1.52 &    14.62 &     0.86 &     2.41
\enddata
\label{agntable}
\end{deluxetable*}

\end{document}